# Surface Acoustic Wave Gas Sensors: Innovations in Functional Materials, Sensing Dynamics, and Signal Analysis

Suman Acharya[a], Balasubramanian Srinivasan[a]*, David Shanahan[b], Utz Roedig[b], Alan O'Riordan[a], Veda Sandeep Nagaraja[a]

[a]*Micro & Nano Systems, Tyndall National Institute, University College Cork, Lee Maltings Complex Dyke Parade, Cork, Cork, T12 R5CP*

[b]*School of Computer Science and Information Technology, University College Cork, Cork, Ireland*

**Abstract**

Surface Acoustic Wave gas sensors have garnered increasing attention as highly sensitive, miniaturized, and wireless-compatible platforms for molecular detection. Their unique ability to convert surface perturbations into measurable acoustic shifts makes them ideal for gas sensing across diverse environments. This review synthesizes reported SAW platforms across substrates and modes (Rayleigh, SH-SAW, Love), links transduction pathways (mass loading, elastic/acoustoelastic, and acoustoelectric effects) to material choice, and benchmarks performance for key analytes (e.g., $NO_2$, $NH_3$, VOCs, $CO_2$, etc). We catalogue nanostructured oxides, polymers, carbon-based films, and hybrid/heterojunction coatings, highlighting attributes such as porosity, surface chemistry, and interfacial charge transfer that govern sensitivity and reversibility. We also highlight the emerging use of SAW devices to probe adsorption–desorption dynamics, offering analyte-specific interaction signatures beyond equilibrium, offering a new perspective into analyte-specific interaction pathways. Additionally, the integration of machine learning is discussed as a transformative tool for signal decoding, environmental compensation, and adaptive calibration. We also identify key challenges: cross-sensitivity, signal drift, material degradation, and deployment at the edge and review recent strategies to address them. Looking ahead, we envision the evolution of SAW platforms into intelligent, autonomous sensing systems with applications in environmental monitoring, industrial process control, and healthcare diagnostics.

**Keywords:** SAW devices, Gas sensors, MEMS sensors, IDTs, Piezoelectric materials, Sensing materials

## 1. Introduction

MicroElectroMechanical Systems (MEMS)-based gas sensors have significantly broadened their application scope in recent years, evolving from traditional industrial and agricultural domains to critical fields such as medical diagnostics, military systems, environmental monitoring, consumer electronics, and homeland security[1–6]. Accurate gas sensing is essential for maintaining safety, controlling industrial processes, and mitigating environmental pollution. MEMS-based gas sensing technologies predominantly utilize optical, resistive, capacitive, and acoustic resonant detection mechanisms[7–9]. While resistive and capacitive sensors offer benefits such as low cost, simple fabrication, and compact size, they often face limitations like limited accuracy, low selectivity, significant nonlinearity, and susceptibility to temperature and humidity fluctuations[2–5]. Optical sensors provide higher precision and low detection limits but are limited by their high cost, large size, complex integration, and substantial power consumption, restricting their adoption in compact or portable devices[10–12].

* **Corresponding Author:** Balasubramanian Srinivasan, Postdoctoral Researcher, Micro & Nano Systems, Tyndall National Institute, University College Cork, Lee Maltings Complex Dyke Parade, Cork, Cork, T12 R5CP
**Email:** bala.srinivasan@tyndall.ie

Surface Acoustic Wave (SAW) sensors, leveraging acoustic waves propagating along the surface of piezoelectric substrates, have emerged as an attractive alternative. SAW was first explained by Lord Rayleigh who, in his classical work, described the surface acoustic modes and their properties. A SAW is an acoustic wave traveling along the surface of an elastic material, characterized by an amplitude that typically decays exponentially with depth into the substrate[13]. SAWs specifically exhibit shallow penetration depths—approximately one acoustic wavelength—resulting in energy being predominantly localized at the substrate's surface. This high surface energy density is a critical factor contributing to the exceptional sensitivity of SAW-based gas sensors. Collectively, SAW sensors uniquely combine ultra-high sensitivity, compact design, flexibility in wired and wireless operations, and seamless integration with Internet of Things (IoT) platforms. SAW devices utilize interdigitated transducers (IDTs)—periodically arranged metallic electrodes deposited onto piezoelectric substrates such as quartz, lithium niobate, or lithium tantalate. Upon application of an alternating voltage, acoustic waves are generated and propagate along the substrate surface. The second IDT captures these waves, converting them back into electrical signals. The presence of target gases is detected via shifts in acoustic wave properties (frequency, amplitude, velocity, or phase) induced by gas molecule interactions with specialized sensing layers deposited atop the substrate[14–16]. Fig. 1 shows a brief description of the SAW working principle, gives a summary of a need for selective and sensitive gas sensors in various fields and how it compares with all the other existing gas sensors.

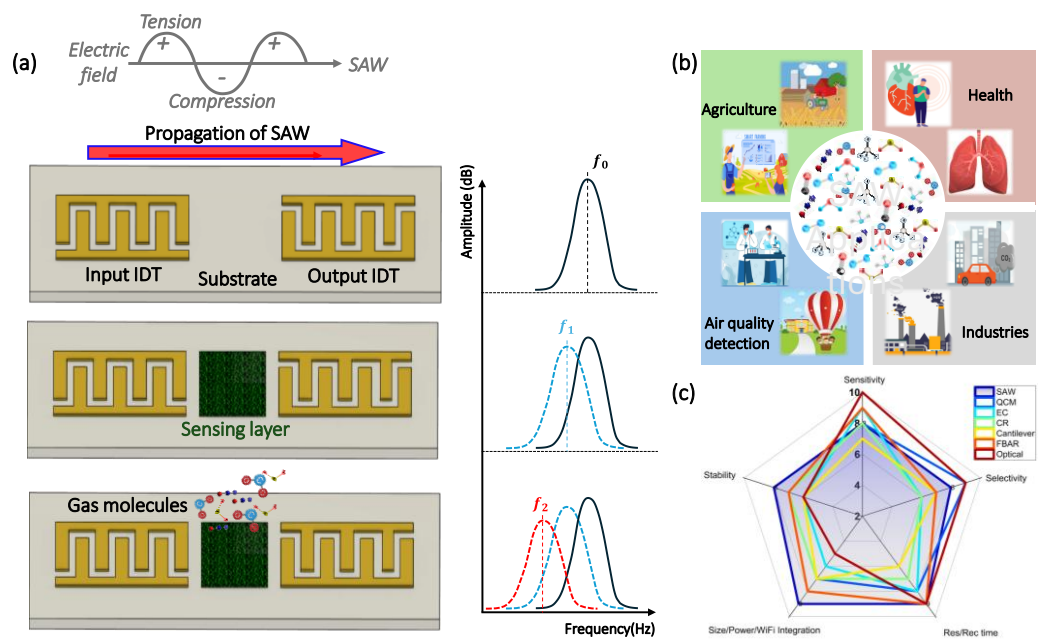

**Fig. 1. (a)** Schematic illustration of the working principle of a surface acoustic wave (SAW) gas sensor. Acoustic waves are generated via the input IDT, propagate along the piezoelectric substrate, and are modulated by a sensing layer. Upon interaction with target gas molecules, changes in mass, elasticity, or damping induce a shift in resonance frequency from the baseline ($f_0$) to $f_1$ (after sensing layer deposition) and to $f_2$ (post gas adsorption). **(b)** Broad application domains of SAW-based gas sensors, including precision agriculture, environmental air quality monitoring, industrial emissions control, and healthcare diagnostics. **(c)** Comparative radar chart showing key performance metrics—sensitivity, selectivity, stability, response/recovery time, and integration potential (size, power, and wireless capability)—of SAW sensors against other conventional gas sensing platforms, highlighting the comprehensive advantages of SAW-based technologies.

Surface Acoustic Wave generation on a piezoelectric material was first demonstrated by White and Voltmer in 1965 by using IDTs. This technology was subsequently adapted for chemical gas detection by Wohltjen and Dessy in 1979, who introduced sensitive coatings for organic gas [1,17–20]. Since then, substantial research efforts have enabled the detection of various gases, including methane ($CH_4$), hydrogen sulfide ($H_2S$), nitrogen dioxide ($NO_2$), carbon dioxide ($CO_2$), ammonia ($NH_3$), hydrogen ($H_2$), and volatile organic compounds (VOCs). The choice of sensing layer materials critically influences the performance of SAW sensors. These materials can be polymeric (e.g., polyvinyl alcohol, polyisobutylene), metal oxides (e.g., ZnO, $SnO_2$), or advanced nanocomposites and nanomaterials such as graphene, carbon nanotubes, and crystalline porous materials such as metal-organic frameworks (MOFs) and covalent organic frameworks (COFs). Each type of material offers distinct affinity characteristics towards specific analytes, thus significantly enhancing sensitivity and selectivity.

SAW sensors offer notable advantages over other acoustic wave-based sensors such as Bulk Acoustic Wave (BAW) like Quartz Crystal Microbalance (QCM). BAW sensors operate with acoustic waves propagating through the bulk of the piezoelectric substrate, which typically makes them less sensitive to surface interactions compared to SAW sensors. They are often preferred for applications requiring higher operating frequencies and robustness in harsh conditions. However, BAW sensors usually exhibit lower sensitivity to surface mass loading due to their bulk mode operation[21–23]. For example, QCM sensors utilize thickness-shear acoustic waves propagating through the bulk of quartz crystals, offering high mass sensitivity but typically at lower operational frequencies[24–26]. While QCM sensors have been widely employed due to their simplicity and cost-effectiveness, their bulk nature limits miniaturization and reduces their compatibility with wireless or passive operation modes. SAW sensors uniquely leverage surface-bound acoustic waves, providing superior surface sensitivity, compactness, and facilitating wireless, passive, and remote sensing applications.

Recent advancements in piezoelectric substrates, including lithium niobate and lithium tantalate, have expanded operational bandwidth and environmental resilience. IDT innovations such as single-phase unidirectional transducers (SPUDTs) and chirped IDTs have significantly improved sensitivity and reduced signal noise. Advanced nanostructured films, including graphene, MXenes, and MOFs, have further enhanced selectivity and reduced detection limits. These technological improvements allow SAW sensors to perform reliably in challenging environments, including high temperatures and corrosive conditions.

Despite substantial advancements, SAW-based gas sensors still face several critical technological challenges that must be addressed for broader commercial adoption. One primary issue is fabrication consistency; achieving reproducible sensor performance across large-scale manufacturing processes remains difficult, particularly for devices relying on advanced nanostructured sensing layers and complex IDT configurations[27]. Long-term stability is another significant concern, as environmental conditions, such as temperature fluctuations, humidity, and mechanical stress, can degrade sensor performance over time. Furthermore, integrating SAW sensors into practical systems poses additional challenges related to packaging, robust signal processing, and ensuring compatibility with existing electronic interfaces. Addressing these challenges through improved fabrication methods, advanced packaging techniques, and robust environmental compensation strategies will be essential for enhancing the reliability and practical utility of SAW-based gas sensing technology. Emerging strategies, such as integrating multi-sensor arrays and employing machine learning and artificial intelligence, are increasingly explored to address accuracy, selectivity, and robustness concerns. Table 1 provides a comparative overview of MEMS-based gas sensor technologies, emphasizing SAW sensors' distinctive blend of advantages, including high sensitivity, operational flexibility, and integrability.

Table 1: Comparative summary of major gas sensing techniques—including capacitive, resistive, optical, and SAW-based sensors—detailing their working mechanisms, key benefits, and limitations, as reported in the literature.[28].

| Sensing technique | Mechanisms | Advantages | Disadvantages |
|---|---|---|---|
| Capacitive | Change in the capacitance due to the change in the permittivity of the sensing film. | Low power consumption, Simple readout circuits, Non-moving structure, | Temperature dependency, non-linear responses |
| Resistive | Change in the resistance of a sensing layer when it encounters the gas molecules. | Simple structure, wide measurement range, low cost | High nonlinearity, poor repeatability, temperature dependency, slower response time |
| Optical | Variation in the refractive index with the absorption of gas molecules. | High sensitivity, thermal stability, low attenuation | High cost, bulky |
| SAW | Change in the frequency/velocity due to change in stiffness or mass of the sensing layer. | High sensitivity, low cost (Expensive in comparison to resistive based), high accuracy, ability to work in wired and wireless mode | High temperature sensitive, sensitive to external vibrations |

While numerous studies have explored individual aspects of SAW-based gas sensors such as piezoelectric substrate materials, IDT configurations, sensing layers, and specific gas detection applications, comprehensive reviews synthesizing these scattered findings into a cohesive framework remain limited. Existing literature often separately addresses these individual components, making it challenging for researchers and practitioners to obtain an integrated understanding of recent advances, emerging trends, and ongoing challenges in SAW sensor technology. This review explicitly addresses this gap by systematically consolidating key research developments across critical domains of SAW gas sensing, including detailed analyses of materials selection, advanced transducer designs, functional sensing layers, and performance metrics. Furthermore, the review critically evaluates the strengths and limitations of different technological approaches and outlines clear future research pathways. By synthesizing dispersed knowledge and highlighting integrative insights, this review aims to serve as a comprehensive resource, facilitating informed research, design decisions, and practical implementation of SAW-based gas sensing technologies.

This review is organized as follows: Section 2 presents an overview of various piezoelectric materials commonly utilized in SAW-based gas sensors, along with their fundamental sensing mechanisms. Section 3 discusses different IDT configurations, emphasizing their design principles and practical applications. In Section 4, we summarize recent advances in functional sensing layers, highlighting the selection criteria and performance characteristics crucial for detecting specific gases. Section 5 provides a comprehensive analysis and comparative data on the performance of SAW-based gas sensors incorporating diverse materials and designs. Finally, Section 6 identifies current technological challenges and proposes recommendations for future development aimed at enhancing the sensitivity, reliability, and practical integration of SAW-based gas sensors.

## 2. Working Principles of SAW devices

SAW sensors function as indirect probes of physical and chemical interactions by exploiting the sensitivity of surface acoustic waves to environmental perturbations. When a chemical entity is present in the propagation path of the SAW, it alters the phase velocity and attenuation of the wave. Such variations originate from the interaction of analytes with a thin sensing layer coated on the piezoelectric substrate. These changes are transduced into electrical signals by IDTs, allowing quantitative detection of the analyte.

Two principal device configurations are employed: resonators and delay line (Fig. 2). A resonator configuration employs IDTs along with grating reflectors that form a cavity for wave resonance (Fig. 2a and 2b). In a two-port resonator, one IDT emits and the other receives the wave, while reflectors create feedback to maintain resonance. This can be also implemented as one-port resonators where single IDT for both purposes (Fig. 2a). Resonators exhibit narrower bandwidths, smaller phase changes ($\sim\pi$), lower insertion losses, and simpler circuit integration. The sensing layer may be deposited within or near the resonating cavity. Despite their structural differences, both configurations operate on the same underlying principle: analyte-induced changes in wave propagation properties.

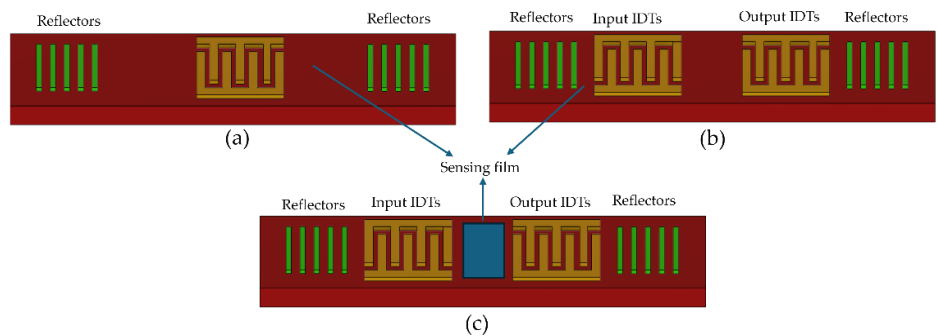

**Fig. 2.** Schematic representations of SAW-based gas sensor configurations: **(a)** single-port resonator, where a single IDT and reflector define the sensing cavity; **(b)** two-port resonator, utilizing separate input and output IDTs with reflectors to form a resonating cavity; and **(c)** two-port delay line, where acoustic waves travel between input and output IDTs across a defined sensing path.

In contrast, two-port delay line (Fig. 2c) comprises two spatially separated IDTs on a piezoelectric substrate-one for excitation and the other for detection. The region between the IDTs is coated with a sensing overlayer, which interacts with target analyte, causing a delay in the transmitted wave proportional to changes in wave velocity. A reflective delay line (one-port) uses a single IDT and a reflector to achieve similar functionality. Delay lines often require precise impedance matching and are prone to larger phase changes ($\sim 2\pi$) making oscillator circuit design more complex, yet they remain practical and effective for sensing applications.

The output signal of a SAW device reflects the velocity ($v$) and attenuation ($\alpha$) changes of the propagating wave, which are typically inferred from shifts in resonance frequency ($f$) or phase ($\phi$). These quantities are related through:

$$\frac{\Delta v}{v_0} = -\frac{\Delta \phi}{\phi_0} = \frac{\Delta f}{f_0} \qquad (1)$$

where $v_0$, $\phi_0$ and $f_0$ denote the unperturbed velocity, phase, and frequency, respectively, and $\Delta$ denotes the corresponding change upon analyte exposure.

The velocity and attenuation shifts originate from multiple interactions between the acoustic wave and the sensing layer. These include mass loading, elastic loading, and acoustoelectric effects, each of which alters the wave's propagation via different mechanisms. The overall perturbation of the wave is described by the complex propagation coefficient $\gamma = \alpha + jk$, where:

$$\frac{\Delta \gamma}{k_0} = \frac{\Delta \alpha}{k_0} - j\frac{\Delta v}{v_0} = \frac{\delta \gamma}{\delta m}\Delta m + \frac{\delta \gamma}{\delta p_{mech}}\Delta p_{mech} + \frac{\delta \gamma}{\delta p_{elec}}\Delta p_{elec} + \frac{\delta \gamma}{\delta p_{env}}\Delta p_{env} \qquad (2)$$

Here, each term corresponds to perturbations from mass, mechanical, electrical, and environmental factors, respectively. For chemically sensitive layers that are acoustically thin, isotropic, and non-piezoelectric, the velocity shift can be analytically decomposed as:

$$\frac{\Delta v}{v_0} = -C_m f_0 h \Delta \rho + C_e f_0 h \left(\frac{4\mu}{v_0^2}\frac{\lambda + \mu}{\lambda + 2\mu}\right) - K^2 \frac{\Delta \sigma^2}{\sigma^2 + v_0^2 C_s^2} \qquad (3)$$

$$\frac{\Delta \alpha}{k} = -K^2 \frac{v_0 C_s \sigma}{\sigma^2 + v_0^2 C_s^2} \qquad (4)$$

Here, $C_m$ and $C_e$ are substrate sensitivity coefficients for mass and elasticity; $h$, $\rho$, $\mu$, $\lambda$, and $\sigma$ are the film thickness, density, shear modulus, bulk modulus, and sheet conductivity, respectively; $C_s$ is the substrate capacitance per unit length. The equations demonstrate that mass loading decreases the wave velocity, elasticity tends to increase it, and conductivity can result in both attenuation and velocity changes, often nonlinearly. Wohltjen et al.[29] first demonstrated frequency shifts induced by mass loading in coated SAW devices. Ricco et al.[30] later showed that the sensor response could be dominated by conductivity changes, particularly with conductive sensing films like lead phthalocyanine. Elastic loading, though historically neglected, was shown by several authors[31,32] and others to significantly affect sensor response in viscoelastic polymer films, particularly due to swelling-induced modulus changes upon vapor exposure.

Further, to evaluate sensor efficacy, three performance metrics are central: selectivity, sensitivity, and stability. Selectivity is determined by the differential response of the sensing layer to target versus non-target species and is often achieved through material design. Sensitivity refers to the smallest detectable analyte concentration and is influenced by layer properties, frequency, and environmental conditions. Stability encompasses both temporal and environmental robustness of the sensor response and is essential for practical applications.

Together, these principles and mechanisms form the foundational basis for designing and analyzing SAW-based chemical sensors with high performance, reliability, and selectivity across diverse sensing environments.

Table 2. Comparative overview of common SAW gas sensor configurations, including resonator and delay line architectures. The table outlines structural features, signal response mechanisms, relative sensitivities, and design considerations relevant to sensor performance and application requirements.

| Configuration | Type | Structure | Signal Response | Sensitivity | Remarks |
|---|---|---|---|---|---|
| Single-Port Resonator | One-port resonator | One IDT with reflectors on either side | Resonant frequency shift | High | Compact design; narrow bandwidth; simpler oscillator circuit; low insertion loss |
| Dual-Port Resonator | Resonator (2-port) | Two IDTs + grating reflectors | Frequency and | Very High | High Q-factor; enhanced selectivity; stable |

| | | forming a resonant cavity | amplitude shift | | response; moderate circuit complexity |
| --- | --- | --- | --- | --- | --- |
| Dual-Port Delay Line | Delay line (2-port) | Input and output IDTs separated by sensing region | Phase shift or frequency shift | Moderate to High | Sensitive to environmental noise; requires impedance matching; large phase shifts (~2π) |

## 3. Piezoelectric Material Substrates for SAW Devices

In SAW-based sensing platforms, surface acoustic waves are launched and detected via interdigitated transducers (IDTs) deposited on a piezoelectric substrate. The propagation behaviour of these waves, specifically their velocity, attenuation, and coupling efficiency, is fundamentally governed by the intrinsic piezoelectric and elastic properties of the substrate material. Piezoelectricity, a property of non-centrosymmetric crystals, allows mechanical strain to generate electric polarization and vice versa. This electromechanical coupling underpins the operation of SAW sensors.

The linear constitutive equations for piezoelectric materials in tensor form describe the relationship between mechanical and electrical quantities as[33]:

$$T = C^E S - eE \tag{5}$$

$$D = \varepsilon^S E + \varepsilon S \tag{6}$$

Here, $T$ and $S$ are the stress and strain tensors, $E$ is the electric field, and $D$ is the electric displacement. The constants $C^E$, $e$, and $\varepsilon^S$ represent the stiffness matrix at constant electric field, the piezoelectric coupling tensor, and the dielectric permittivity at constant strain, respectively.

The choice of substrate material is critical and directly impacts device performance. Several parameters should be considered in substrate selection:

1. **Crystal Cut and Orientation:** Acoustic wave velocity and polarization depend heavily on the crystallographic orientation. Specific cuts like the 128° YX-cut in lithium niobate (LiNbO₃) are widely used due to high electromechanical coupling and stable Rayleigh wave propagation.
2. **Electromechanical Coupling Coefficient ($k^2$):** A higher coupling coefficient enhances the transduction efficiency between electrical and mechanical energy, improving sensor sensitivity and bandwidth.
3. **Acoustic Wave Velocity:** High velocity substrates support higher frequency operations, which is favourable for sensitive detection.
4. **Thermal Stability:** Substrates with low temperature coefficients of frequency (TCF) minimize frequency drift under thermal variations, ensuring long-term reliability.
5. **Sensing Layer Compatibility:** Chemical adhesion, lattice compatibility, and thermal matching between the substrate and the sensing layer influence stability and durability.
6. **Mechanical Robustness and Fabrication Compatibility:** For integration in microsystems, substrates must support thin-film deposition, etching, and CMOS-compatible processes[34].

Among all substrates used in SAW devices, **Lithium Niobate (LiNbO₃)** remains a leading material due to its high electromechanical coupling coefficient ($k^2$)—around 5.4% for the widely used 128° YX-cut—which promotes efficient energy transduction and high sensitivity. It supports strong Rayleigh and shear horizontal (SH) waves, making it versatile for gas and biosensing applications. However, a key limitation is its relatively high **temperature coefficient of frequency (TCF)** (~90 ppm/°C), which causes frequency drift in variable thermal environments and necessitates thermal compensation strategies in precision applications[35,36]

**Quartz (SiO$_2$)**, particularly in the ST-cut orientation, offers superior thermal stability (TCF ≈ 0 ppm/°C) and low acoustic loss, making it the gold standard for frequency-stable reference devices and delay lines. Its major drawback is the very low $k^2$ (~0.1%), limiting its sensitivity in chemical sensing. Nevertheless, quartz is still used in SAW sensors where temperature invariance and signal stability outweigh raw sensitivity, such as in environmental monitors and frequency-control circuits[37–39].

**Zinc Oxide (ZnO)** thin films are increasingly employed in SAW devices for their relatively high $k^2$ (1.5–3%), ease of deposition on various substrates, and capability to support multiple acoustic modes (Rayleigh, Love, and SH waves). ZnO is also compatible with flexible and stretchable substrates, making it suitable for wearable sensors and implantable biomedical devices. However, its poor chemical stability in acidic or humid environments necessitates protective coatings or encapsulation[39–41].

**Aluminum Nitride (AlN)** has emerged as a top choice for RF SAW filters and high-frequency sensors. With bulk acoustic wave velocities reaching ~11,000 m/s and excellent thermal conductivity, thin-film AlN supports GHz-level operation and is inherently stable in high-temperature environments. Its moderate $k^2$ (~0.7–1.5%) is enhanced through **scandium doping (ScAlN)**, which boosts piezoelectric response without sacrificing thermal robustness. Moreover, AlN is fully compatible with CMOS processing, enabling monolithic integration of SAW sensors with microelectronics[42–44].

**Lithium Tantalate (LiTaO$_3$)** is a strong alternative to LiNbO$_3$, offering slightly lower $k^2$ (≈4.3%) but significantly improved thermal stability (TCF ≈ 30–35 ppm/°C). Its robust dielectric properties make it suitable for high-temperature sensing in harsh environments such as industrial exhaust or aerospace systems. Notably, 36° YX-cut LiTaO$_3$ supports efficient SH-SAW propagation, further expanding its application in humidity and biosensors[45,46].

Advanced multilayer substrates and thin-film materials like **PZT (lead zirconate titanate)** offer superior piezoelectric coefficients and tunability. Randomly oriented PZT films demonstrate $k^2$ exceeding 10%, enabling highly miniaturized and sensitive SAW resonators. However, PZT's chemical complexity, thermal mismatch with silicon, and integration challenges limit its widespread adoption. It has shown promise in biomedical sensing when fabricated on compliant polymeric or glass substrates[47,48].

Moreover, **hybrid or bonded substrates**—such as **LiNbO$_3$-on-quartz** or **LiTaO$_3$-on-quartz**—combine the high $k^2$ of active layers with the thermal stability of quartz. This layered approach enables both high sensitivity and low thermal drift, especially advantageous for precision sensors operating across varying environmental conditions[35,49]. The table 3 below consolidates the critical properties of these substrates for side-by-side evaluation

**Table 3:** Physical and piezoelectric properties of widely used SAW substrate materials—such as LiNbO$_3$, LiTaO$_3$, Quartz, AlN, and LGS—summarized by orientation, dielectric constants, thermal expansion, acoustic velocity, and coupling coefficients, as reported in the literature.

| Material | Orientation | Density (g/cm$^3$) | Poisson's Ratio | Dielectric Constant ($\varepsilon_r$) | CTE (×10$^{-6}$/°C) | Wave Velocity (m/s) | Electromechanical Coupling $k^2$ (%) | Curie Temperature (°C) | Ref |
|---|---|---|---|---|---|---|---|---|---|

| Material | Cut | Density (g/cm³) | Poisson's ratio | Dielectric constant | $k^2$ (%) | Velocity (m/s) | TCF (ppm/°C) | Curie temp (°C) | Ref |
|---|---|---|---|---|---|---|---|---|---|
| LiNbO₃ | Y, Z axis | 4.64 | 0.24 | 85 | 15 | 3488 | 0.045 | 1150 | [18] |
| LiNbO₃ | 128° YX | 4.64 | 0.24 | 85 | 15 | 3992 | 5-11.3 | 1150 | [19] |
| LiNbO₃ | 64° YX | 4.64 | 0.24 | 85 | 15 | 4742 | 11.3 | 1150 | [50] |
| LiNbO₃ | 41° YX | 4.64 | 0.24 | 85 | 15 | 3940 | 17.2 | 1150 | [51] |
| Quartz (SiO₂) | ST-cut | 2.65 | 0.17 | 4.3 | 0 | 3159 | 0.03 | 573 | [52,53] |
| Quartz (SiO₂) | X-cut | 2.65 | 0.17 | 4.3 | 0 | 3159 | 0.14 | 573 | [54] |
| LiTaO₃ | Y, Z cut | 7.45 | 0.23 | 43-54 | 16.5 | 3230 | 0.66 | 607 | [34] |
| LiTaO₃ | X-cut | 7.45 | 0.23 | 43-54 | 16.5 | 3290 | 0.75 | 607 | [34] |
| GaN | - | 6.095 | 0.183 | - | 3.17 | 8000 | 0.13 | - | [55] |
| AlN | - | 3.26 | 0.22 | 8.5-10 | 5.2 | 5800 | 0.7-1.5 | 2200 | [56] |
| LGS | - | 5.74 | 0.32 | 8.16 | - | 2742 | 0.32 | 1470 | [57] |

Material selection for SAW sensors involves a complex optimization of multiple competing parameters. High electromechanical coupling coefficients ($k^2$), as seen in lithium niobate (LiNbO₃), are desirable for maximizing sensitivity and minimizing insertion loss; however, such materials often exhibit high temperature coefficients of frequency (TCF), leading to frequency drift under thermal fluctuations. In contrast, substrates like quartz (SiO₂) offer exceptional thermal stability (near-zero TCF), making them ideal for frequency-stable devices, albeit at the cost of lower sensitivity due to weak piezoelectric coupling. Another important trade-off is between acoustic wave velocity and integration potential. High-velocity materials such as aluminum nitride (AlN) and gallium nitride (GaN) enable operation at GHz frequencies and facilitate device miniaturization, but issues such as thin-film stress, lattice mismatch with silicon, and limited coupling in undoped films must be addressed. Similarly, thin-film ZnO offers good coupling and flexibility, making it attractive for wearable applications, though its chemical vulnerability necessitates protective coatings. Table 4 gives comprehensive details on how the piezoelectric materials are applied for various SAW sensor applications in the existing literatures.

Emerging strategies, such as hybrid substrates (e.g., LiNbO₃-on-quartz, AlN-on-silicon), seek to combine the advantages of different materials—pairing high sensitivity with thermal stability or mechanical robustness. However, these approaches introduce fabrication complexity and potential adhesion challenges. Consequently, substrate choice is inherently application-specific, requiring careful balancing of sensitivity, thermal stability, manufacturability, and environmental resilience to meet operational demands.

Table 4. Selection of piezoelectric substrates for various SAW sensor application domains, based on their mechanical, electrical, and integration advantages, as reported in the literature.

| Application Focus | Preferred Substrate | Key Reason |
|---|---|---|
| High sensitivity and miniaturization | LiNbO$_3$, PZT | High electromechanical coupling coefficient |
| Temperature-invariant sensing | ST-cut Quartz | Near-zero TCF for thermal stability |
| High-frequency RF operation | AlN, ScAlN | High acoustic velocity, GHz operation |
| Flexible and wearable devices | ZnO thin films | Flexibility, ease of thin-film deposition |
| CMOS integration and harsh environments | AlN, GaN | High-temperature stability, silicon compatibility |

## 3.1. Types of Acoustic Waves in SAW Devices and Material Considerations

Numerous SAW devices have been developed, each designed to operate at ultrasonic frequencies where acoustic waves propagate along or near the surface of a piezoelectric substrate. As discussed in the previous section, the working principle of SAW devices fundamentally relies on the nature of wave propagation within the substrate material. Different types of acoustic waves can be generated depending on substrate properties and device structure, and each wave type offers distinct advantages for sensing applications. Particle motion direction, penetration depth, wave velocity, and energy confinement are critical characteristics differentiating these wave types. The primary acoustic wave modes utilized in SAW devices are Rayleigh waves, Lamb waves, Love waves, and Shear Horizontal (SH) waves, which are discussed in detail below.

### 3.1.1. Rayleigh Waves

Rayleigh waves are classical surface acoustic waves that propagate along the substrate surface. First described by Lord Rayleigh in 1885, these waves involve elliptical particle motion in the vertical plane, combining vertical and longitudinal displacements. The amplitude of the wave decays exponentially with depth, typically penetrating to about one acoustic wavelength beneath the surface[13]. The wave velocity $C_R$ of Rayleigh waves is related to the shear wave velocity $C_S$ and Poisson's ratio $v$ of the substrate by the expression:

$$C_R = C_s \left( \frac{0.87 + 1.12v}{1 + v} \right) \qquad (7)$$

Piezoelectric materials such as 128° YX-cut LiNbO$_3$, ST-cut quartz, and 41° YX-cut LiNbO$_3$ are commonly used to generate Rayleigh waves. These waves require relatively low power to excite, are cost-effective to manufacture, and are highly suitable for gas-phase sensing due to their strong surface sensitivity[18–20]. However, their vertical displacement component leads to significant energy loss in liquid environments, making them less ideal for fluidic sensing applications (Table 5). Rayleigh waves typically propagate with velocities between 3000 and 4000 m/s depending on the substrate, and they exhibit slow propagation speeds, approximately 10$^{-5}$ times the speed of light. They operate across a frequency range from 3 MHz to 2 GHz, although most gas sensors are optimized between 50 MHz and 350 MHz to balance sensitivity and minimize noise. Recent studies have demonstrated that Rayleigh wave SAW devices functionalized with two-dimensional (2D) materials such as graphene and MoS$_2$, or with metal-organic frameworks (MOFs), offer enhanced detection sensitivity for volatile organic compounds and toxic gases at parts-per-billion levels.

### 3.1.2. Lamb Waves

Lamb waves are guided elastic waves that propagate within thin-walled structures, bounded by the top and bottom surfaces of a substrate. They exhibit both normal and shear motion components and are highly sensitive to surface perturbations. The velocity of Lamb waves strongly depends on the frequency-thickness product of the substrate, governed by the dispersion relationship for Lamb modes. In simplified form for symmetric modes, the phase velocity $C_L$ can be approximated by:

$$C_L = \sqrt{\frac{E}{\rho(1-\nu^2)}} \tag{8}$$

where $E$ is Young's modulus, $\rho$ is the density, and $\nu$ is Poisson's ratio of the material. The penetration depth of Lamb waves is approximately equal to the thickness of the thin substrate layer. In natural crystalline substrates, Lamb waves operate between 200 kHz and 2 MHz, while in thin-film engineered devices such as ZnO, AlN, or PVDF films, operating frequencies extend from 200 MHz up to 2 GHz[34,55,58,59]. Lamb wave propagation velocities are highly dispersive, typically ranging between 1000 and 4000 m/s depending on substrate thickness and excitation frequency. Due to their low phase velocities and minimal acoustic streaming effects, Lamb waves are highly favourable for sensing applications in liquids where energy leakage must be minimized[60] (Table 5). Recent developments[61–63] have enhanced Lamb wave sensitivity using multilayer ZnO-AlN composites for biochemical and humidity sensing.

### 3.1.3. Love Waves

Love waves are horizontally polarized shear waves guided within a thin deposited guiding layer over a piezoelectric substrate. These waves propagate through a two-layer structure, where a thin elastic layer, typically SiO$_2$, ZnO, or TiO$_2$, is deposited atop a thicker substrate such as 36° YX-cut lithium tantalate (LiTaO$_3$) or 64° YX-cut lithium niobate (LiNbO$_3$)[34,64–67]. The particle motion is parallel to the substrate surface and orthogonal to the direction of wave propagation. The energy of Love waves is tightly confined within the guiding layer, resulting in enhanced sensitivity and reduced energy leakage into adjacent media. The phase velocity $C_{Love}$ of Love waves can be approximated by:

$$C_{Love} = C_S \sqrt{1 - \frac{2\rho_g d}{\rho_s \lambda}} \tag{9}$$

where $C_S$ is the shear velocity of the substrate, $\rho_g$ and $\rho_s$ are the densities of the guiding and substrate layers respectively, $d$ is the guiding layer thickness, and $\lambda$ is the wavelength. Love waves typically propagate with velocities between 2000 and 3500 m/s depending on the substrate and guiding layer properties, with operational frequency ranges between 100 MHz and 450 MHz. Due to their strong confinement and minimal energy radiation into liquids, Love waves are highly suitable for biosensing and liquid-phase applications (Table 5). Recent studies by Wang et al. and Chen et al. have demonstrated the fabrication of Love wave sensors functionalized with hybrid guiding layers composed of SiO$_2$ and advanced nanomaterials, achieving ultrahigh sensitivity in detecting biological species, pathogens, and chemical markers.

### 3.1.4. Shear Horizontal (SH) Waves

Shear Horizontal (SH) waves propagate with particle displacement entirely confined to the plane of the substrate surface and perpendicular to the wave propagation direction. Unlike Rayleigh waves, SH waves maintain strong surface sensitivity without exponential amplitude decay into the substrate. The phase velocity $C_{SH}$ for SH waves is closely approximated by:

$$C_{SH} = \sqrt{\frac{G}{\rho}} \tag{10}$$

where $G$ is the shear modulus and $\rho$ is the density of the substrate material. SH waves are typically excited in substrates such as 36° YX-cut $LiTaO_3$, 34° YX-cut $LiNbO_3$, 64° YX-cut $LiNbO_3$, and quartz[34,68–71]. They propagate with velocities ranging between 2500 and 3800 m/s depending on the material system. SH-SAW devices generally operate at frequencies greater than 100 MHz and offer excellent mechanical robustness, high sensitivity, and low noise characteristics, making them ideal for biochemical sensing and gas detection even in humid environments (Table 5). Recent reports have demonstrated the use of SH-SAW devices functionalized with nanostructured MOFs, resulting in highly selective detection of hazardous gases in complex environmental conditions[72].

Table 5. Classification of surface acoustic wave types used in SAW-based sensors, detailing their wave motion, penetration characteristics, compatible media, substrate materials, and sensing applications, as reported in the literature.

| Wave Type | Particle Motion | Penetration Depth | Suitable Medium | Materials/Substrates | Applications | REF |
|---|---|---|---|---|---|---|
| Rayleigh | Elliptical (vertical + longitudinal) | ~1 acoustic wavelength | Gas-phase | 128° YX $LiNbO_3$, Quartz | VOC sensing, toxic gas detection | [13,19,20] |
| Lamb | Mixed (vertical + shear) | Thickness of thin film | Gas/Liquid | ZnO/AlN thin films | Humidity sensing, liquid sensing | [55,58,59] |
| Love | Horizontal (within guiding layer) | Guiding layer thickness | Liquid-phase | $LiTaO_3$ + $SiO_2$, ZnO | Biosensing, virus detection | [34,64–67] |
| Shear Horizontal (SH) | Shear-horizontal (surface plane) | Surface-confined | Gas/Liquid | $LiNbO_3$, $LiTaO_3$, Quartz | Biochemical sensing, wearable sensors | [34,68–71] |

## 3.2. Configuration of the Interdigitated Electrodes in the SAW devices

The excitation and detection of SAWs in piezoelectric substrates are primarily achieved using IDTs. When an alternating current signal is applied across the electrodes of an IDT, periodic mechanical deformations are induced in the piezoelectric substrate, resulting in the generation of propagating acoustic waves. Among the many methods available for generating SAWs, IDTs offer a compact, efficient, and CMOS-compatible approach, making them indispensable for gas sensing, biochemical detection, RF filtering, and microfluidic applications.

The performance of a SAW device is highly sensitive to the design of the IDTs, and critical factors such as acoustic wavelength ($\lambda$), acoustic velocity ($v$), operational frequency ($f = v/\lambda$), aperture, number of finger pairs, electrode thickness, metallization ratio (ratio of electrode width to period), and inter-electrode spacing must be optimized for the target application. Various IDT architectures have been developed to enhance specific parameters including electromechanical coupling efficiency, spurious signal suppression, insertion loss reduction, frequency selectivity, and mode purity. The key configurations of IDTs used in SAW devices are described below.

### 3.2.1. Conventional Electrode Configuration

The conventional IDT configuration consists of two symmetric sets of metallic electrodes patterned on the piezoelectric surface, with alternating fingers connected to opposite terminals. Upon the application of an AC voltage, these electrodes generate alternating mechanical strains, launching SAWs at the designed centre frequency corresponding to the electrode periodicity (Fig 3a). The wave propagates through the substrate and is detected by a second (receiving) IDT, where it is converted back into an electrical signal. This simple architecture is widely used due to its ease of fabrication, reliability, and suitability for applications such as RF filters and gas sensors[18,34,73–75]. However, conventional IDTs can exhibit undesired spurious modes, reflections, and harmonic generation if not carefully designed.

### 3.2.2. Single Phase Unidirectional Transducer (SPUDT) Configuration

Single Phase Unidirectional Transducers (SPUDTs) incorporate internal reflectors within the electrode structure to direct acoustic wave propagation preferentially in one direction. By adjusting the electrode widths and spacings (typically using $\lambda/4$ and $\lambda/8$ intervals), the reflected waves are phase-matched and combined constructively on one side while destructively cancelled on the other (Fig 3b). This configuration effectively suppresses backward-traveling waves, thereby reducing insertion loss and triple-transit echoes. SPUDTs are especially useful in wireless SAW sensors, microfluidic actuation, and low-loss communication systems[76–78].

### 3.2.3. Split Electrode Configuration

In split IDTs as shown in Fig. 3c, each electrode is subdivided into narrower segments spaced by $\lambda/8$. This design enables precise control over frequency response and harmonic suppression, allowing the device to operate efficiently at higher-order harmonics (such as the third harmonic) without significantly shrinking the electrode width. Split configurations also offer improved out-of-band rejection and spectral purity, making them attractive for both high-frequency RF applications and highly selective sensing platforms[34,79].

### 3.2.4. Apodized IDT Configuration

Apodized IDTs modify the amplitude of the generated acoustic wave by varying the overlap or width of electrodes along the transducer length as shown in Fig. 3d. Non-parallel or gradually tapered electrode profiles introduce amplitude modulation of the wave, significantly reducing side lobes and suppressing spurious signals. These IDTs provide improved main-lobe-to-sidelobe ratios and broadened bandwidths, which are critical in high-precision RF filters and sensing applications demanding minimal cross-talk and signal distortion[80–82].

### 3.2.5. Floating Electrode Unidirectional Transducer (FEUDT) Configuration

Floating electrode configurations feature unconnected (floating) electrodes inserted between driven electrode fingers (Fig. 3e). The introduction of floating fingers modifies the electrical loading and surface potential distribution, improving unidirectionality and allowing operation at higher centre frequencies for a given electrode width. FEUDTs offer superior insertion loss characteristics compared to conventional IDTs and are advantageous for high-frequency RF filters and high-sensitivity sensor arrays[83–86].

### 3.2.6. Focused IDT Configuration

Focused or circular IDTs arrange electrode fingers along concentric arcs with varying radii but constant width, such that acoustic energy from each finger converges at a geometric focal point. The degree of arc, number of pairs, and focal length must be optimized to maximize wave focusing as shown in Fig. 3f. Focused IDTs generate localized regions of high acoustic intensity, enabling applications such as droplet manipulation in microfluidics, localized biosensing, cell lysis, and quantum acoustics experiments[87–89].

### 3.2.7. Dummy Electrode Configuration

Dummy electrode IDTs introduce additional floating or inactive fingers between active electrodes, serving to balance mass loading, minimize parasitic reflections, and suppress unwanted acoustic modes (Fig. 3g). Dummy fingers improve impedance matching and reduce fabrication-induced asymmetries, leading to enhanced spectral purity and sensitivity. Such configurations are especially common in highly selective RF SAW filters and resonators[90–92].

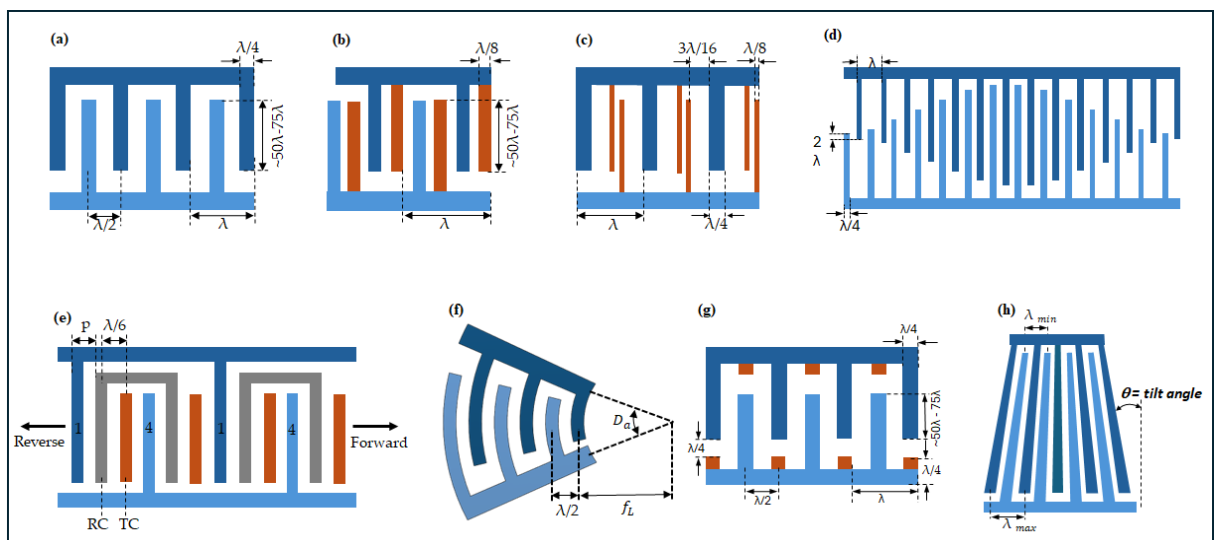

**Fig. 3.** Schematic illustration of various interdigital transducer (IDT) configurations used in SAW devices to optimize wave generation, frequency response, and sensor performance: **(a) Conventional IDT** – standard finger configuration with equal finger widths and spacing [90–92]; **(b) Single-Phase Unidirectional Transducer (SPUDT)** – incorporates phase-shifting electrodes to improve directionality [90–92]; **(c) Split-Finger IDT** – enhances harmonic suppression and bandwidth by splitting electrode fingers with varied spacing [90–92]; **(d) Apodized IDT** – finger overlap tapers to reduce sidelobes and improve frequency selectivity [90–92]; **(e) Floating Electrode Unidirectional Transducer (FEUDT) IDT** – includes floating electrodes between active fingers to modify wave reflection and coupling [90–92]; **(f) Focused IDT** – curved electrode geometry focuses acoustic waves to a point for increased energy concentration [90–92]; **(g) Dummy IDT** – non-functional fingers included to minimize diffraction effects and improve acoustic uniformity [90–92]; **(h) Slanted Finger IDT (SFIT)** – varying finger pitch and tilt angle used for wideband operation and filtering applications [90–92].

### 3.2.8. Slanted (Tapered) IDT Configuration

Slanted IDTs vary the spacing between adjacent electrodes across the aperture width, allowing multiple wavelengths to be excited simultaneously, as shown in Fig. 3h. This leads to wideband frequency responses and reduced reflection artifacts. Slanted configurations are often used in wideband RF devices, broadband SAW sensors, and microfluidic devices requiring frequency agility. By designing appropriate tapers, slanted IDTs can effectively suppress spurious resonances and broaden the operational bandwidth[93–97].

## 4. SAW Sensor Materials, Mechanisms, and Performance

In the development of SAW-based gas sensors, the choice of piezoelectric substrate is primarily governed by the specific requirements of the application, such as operating temperature, stability, and wave propagation characteristics[13,38,98–102]. The geometry and configuration of the IDTs are equally crucial, as they directly influence signal strength, quality factor, and device resolution[60,103]. However, once the IDT design reaches a performance plateau, further improvements in sensing performance must come from optimizing the sensing material[28,104,105]. The sensing layer, deposited over the piezoelectric surface, serves as the active interface where adsorption of gas molecules leads to perturbations in the propagation of surface acoustic waves. These perturbations stem from mass loading, elastic loading, and acoustoelastic effects, ultimately causing shifts in wave velocity and resonant frequency. The frequency shift $\Delta f$ relative to the unperturbed value $f_0$ can be described by the relation[15,29,30,106–110]

$$\frac{\Delta f}{f_0} \cong \frac{\Delta v}{v_0} = -c_m f_0 \Delta\left(\frac{m}{A}\right) + 4c_e \frac{f_0}{v_0^2}\Delta(hG') - \frac{K^2}{2}\Delta\left[\frac{\sigma_s^2}{\sigma_s^2 + v_0^2 C_s^2}\right] \quad (11)$$

where $f_0$ and $v_0$ are the center frequency and unperturbed velocity, $c_m$ and $c_e$ are mass and elastic sensitivity coefficients, $m/A$ is the mass change per unit area, $h$ is the sensing film thickness, $G'$ is the storage modulus, $K^2$ is the electromechanical coupling coefficient, $\sigma_s$ is the sheet conductivity, and $C_s$ is the capacitance per unit length of the piezoelectric substrate.

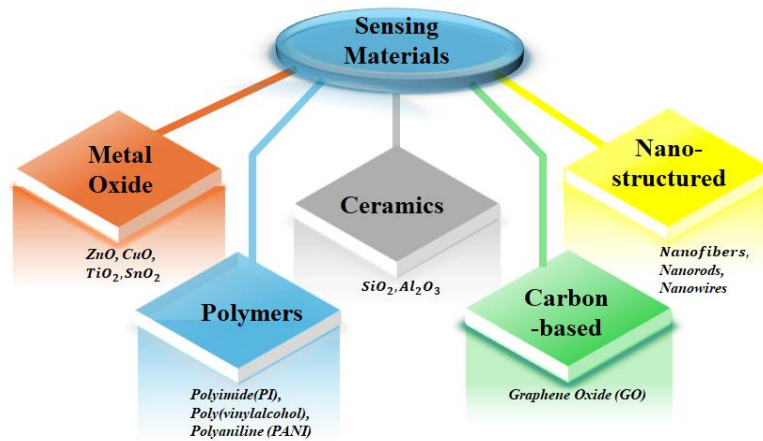

**Fig. 4**. Classification of sensing materials commonly used in gas sensor development. The categories include: **metal oxides** (e.g., ZnO, CuO, TiO$_2$, SnO$_2$), **ceramics** (e.g., SiO$_2$, Al$_2$O$_3$), **polymers** (e.g., polyimide, polyvinyl alcohol, polyaniline), **carbon-based materials** (e.g., graphene oxide), and **nanostructured forms** (e.g., MOFs, nanofibers, nanorods, nanowires). Each class offers distinct physicochemical properties tailored for specific gas adsorption and signal transduction mechanisms in sensor applications.

A wide range of sensing materials have been explored for SAW applications, including metal oxides (e.g., ZnO, CuO, $SnO_2$), ceramics (e.g., $Al_2O_3$, $SiO_2$), polymers (e.g., polyimide, PANI), carbon-based materials (e.g., graphene oxide), and emerging nanostructured hybrids such as MOFs and MXenes (Fig. 4). The performance of the sensing layer is governed by its ability to offer high selectivity, strong adsorption capacity, excellent substrate adhesion, fast and reversible response, environmental stability, and reproducibility[28,105,106,111,112]. Material properties such as grain size, porosity, surface roughness, and deposition uniformity critically influence gas transport and interaction at the sensor surface. For example, InOx thin films fabricated by RF diode sputtering exhibited finer grains and a higher oxygen vacancy density than those made via magnetron sputtering, resulting in superior hydrogen sensitivity ($\sim 11.83\ kHz\ at\ 400\ ppm\ H_2$)[19]. Similarly, $ZnO$–$Al_2O_3$ nanocomposites have been reported to undergo sulfurization to ZnS upon $H_2S$ exposure, yielding a positive frequency shift ($\sim 500\ Hz\ at\ 10\ ppb$) due to elastic loading[53]. Similarly, Cu-doped $SnO_2$ enabled hydrogen detection with sensitivities around $0.069\ kHz/ppm$ and rapid dynamics[113]. Ag nanowire-decorated $SnO_2$/CuO heterostructures detected $SO_2$ at 80°C with sensitivity of $3.696\ kHz/ppm$ and response/recovery times of $40/60\ s$, respectively[114].

The choice of sensing material remains central to achieving high SAW sensor performance. Fig. 5 presents representative examples. CuO nanoparticles (Fig. 5a), as demonstrated by Bak et al., offer reliable $NH_3$ sensitivity at room temperature due to efficient interfacial charge transfer at the CuO/AlN heterojunction. Polyaniline (PANI) nanotubes (Fig. 5b), prepared by interfacial polymerization, exhibit excellent $NH_3$ and VOC detection capabilities due to their conjugated π-system, tunable redox states, and reversible protonation–deprotonation mechanisms. Their tubular morphology further enhances adsorption by increasing the surface-to-volume ratio. Indium oxide films fabricated using glancing angle deposition (GLAD) (Fig. 5c) provide high surface roughness and crystalline orientation. When decorated with plasmonic silver nanoparticles, these films leverage local surface plasmon resonance to enable low-power $NO_2$ sensing via enhanced photocarrier generation and surface charge modulation. $In_2O_3$/$Co_3O_4$ nanocomposites (Fig. 5d) form heterojunctions that enhance charge separation, facilitating $NO_2$ sensing with improved sensitivity. Hierarchical NiO microstructures (Fig. 5e) provide a flower-like morphology with high surface area and porous architecture, beneficial for diffusion-limited gas sensing processes. MOFs, such as $NH_2$-MIL-125(Ti) (Fig. 5f), offer crystalline porous networks with tunable chemical functionality, high surface area, and well-defined adsorption channels[115–119]. These features are critical for selective molecular recognition through adsorption kinetics, making MOFs highly suitable for VOC and other gas detection in SAW platforms. Notably, all three Mg-, Zr- and Ti-based MOFs have shown promise due to their chemical stability and ability to retain crystallinity under operational conditions. These materials exhibit tailored nanostructures—such as porous frameworks (Fig. 5a, 5f), tubular or columnar assemblies (Fig. 5b, 5c), and flower-like architectures (Fig. 5d, 5e)—which significantly enhance gas adsorption via increased surface area and abundance of active sites.

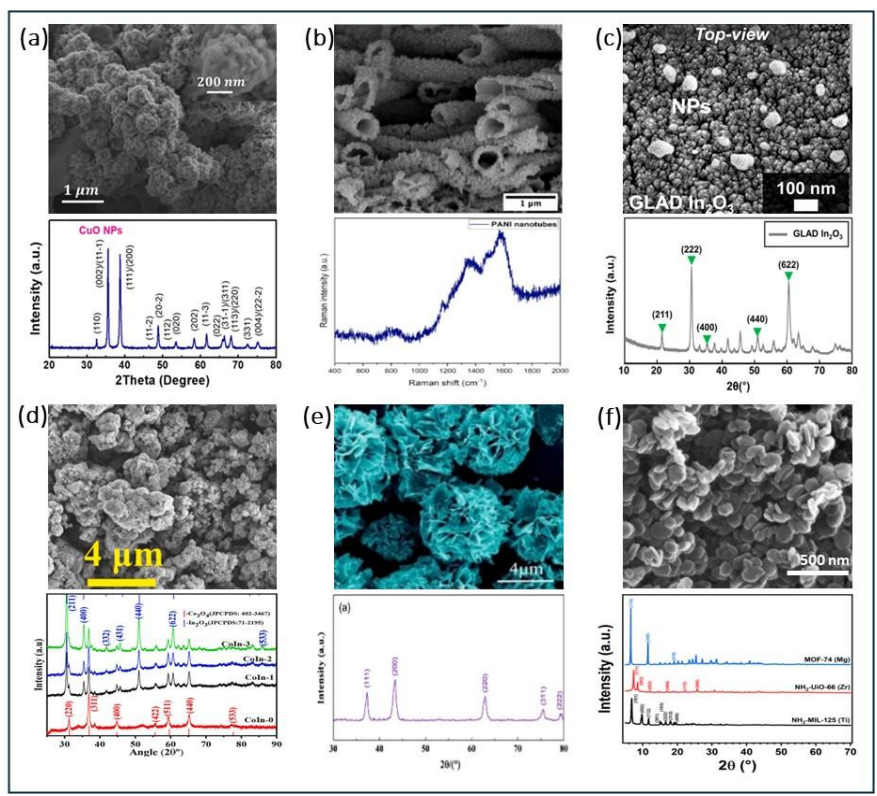

Fig. 5. Representative examples of sensing materials employed in gas sensor design. (a) SEM image and XRD pattern of CuO nanoparticles used in CuO/AlN heterostructure sensors. *Reproduced with permission from*[56] (b) SEM image and Raman spectra of polyaniline (PANI) hollow nanotubes for $H_2$ detection. *Reproduced with permission from*[120] (c) SEM image and XRD pattern of Ag nanoparticle-coated $In_2O_3$ films for photoactivated $NO_2$ sensing. *Reproduced with permission from*[121] (d) SEM image and XRD pattern of $In_2O_3/Co_3O_4$ nanoflowers with enhanced heterojunction interfaces. *Reproduced with permission from*[122] (e) SEM image and XRD pattern of flower-like NiO hierarchical structures used in p-type sensing platforms. *Reproduced with permission from*[123] (f) SEM image and XRD pattern of $NH_2$-MIL-125(Ti) MOFs illustrating high crystallinity and porosity for selective gas adsorption. *Reproduced with permission from*[124]

Recent advances have also highlighted the potential of hybrid nanostructures. ZnO@MXene heterostructures combine the chemical reactivity of ZnO with the high conductivity and surface functional groups of MXenes, resulting in synergistic enhancements in charge transfer, surface reactivity, and response dynamics[125]. Under UV activation, these materials enable sub-ppb level detection of $NH_3$, with fast response and recovery times. Each of these materials presents specific advantages tailored to target gas types, interaction kinetics, and desired sensor operating conditions. As the structural, morphological, and compositional characteristics shown in Fig. 5 collectively demonstrate, rational materials engineering remains pivotal to the continued development of high-performance SAW sensors for real-world applications in environmental monitoring, healthcare, and industrial safety.

### 4.1. Functional Materials, Nanostructures, and Mechanisms Governing SAW Sensor Response

The integration of carbonaceous nanomaterials has substantially expanded the sensing capabilities of SAW platforms across a wide range of target analytes. Zhou et al.[106] utilized suspended graphene/$SiO_2$ structures to achieve parts-per-trillion (ng/L) level acetone detection. GO-PEDOT:PSS hybrids demonstrated $NO_2$ sensitivity of $5.89\,kHz/100\,ppm$ ($31.3\,Hz\,per\,mg/m^3$) with a rapid recovery time of just 10 s[57]. Similarly, porous graphene/PVDF composites enabled fast DMMP detection with a $4.5\,s$ response time[126]. Among polymer-based materials, Teflon AF 2400 facilitated $CO_2$ detection with a rapid recovery time of ~20 $s$ [127] ,while fluorinated polyimide films exhibited humidity sensitivity of ~4.15 $kHz/\%RH$ and a response time around 7 s[128]. Cellulose nanocrystals also showed promise, enabling HCl detection with ~2 $kHz/ppm$ (1340 $Hz\,per\,mg/m^3$) sensitivity at room temperature[129].

Humidity sensing has received growing attention across sectors such as agriculture, healthcare, and environmental monitoring[104,128]. For example, Le et al.[104] fabricated uniform GO films ($thickness, t \approx 210\,nm$) on AlN substrates operating at a resonant frequency, $f_r$ of 226.3 MHz. Their SAW-based humidity sensors achieved response times below 10 $s$ and sensitivity of 111.7/%RH across a 10–90% RH range (Fig. 6a). The use of surface tension-assisted deposition enabled precise control over film thickness. In a complementary effort, Ren et al.[130] developed real-time compensation algorithms for hydrogen SAW sensors, reducing baseline drift by 97.86% and lowering the concentration prediction error from 22.58% to 4.83%.

Expanding the functional domain of SAW platforms, Lim et al. demonstrated dual-layer sensors for simultaneous $CO_2$ and $NO_2$ detection using Teflon AF 2400 and indium tin oxide films on 41° YX $LiNbO_3$ substrates. Although the system exhibited good linearity, $NO_2$ recovery times remained long (~15 $minutes$). Pasupuleti et al.[131] introduced high-performance Langasite-based SAW sensors employing 2D g-$C_3N_4$@$TiO_2$ nanoplates ($f_r = 135.73\,MHz, t \approx 310 \pm 15\,nm$). These sensors showed excellent $NO_2$ sensitivity (~19.7 $kHz/100\,ppm$ (105 $Hz\,per\,mg/m^3$)), a theoretical LOD of ~152 $ppb$ (286 $\mu g/m^3$), and response/recovery times of 143/114 $s$ at room temperature. (Fig. 6b) In another study, the same authors[125] incorporated $Ti_3C_2T_x$ MXene sheets to enhance $NH_3$ sensing capability through improved charge transport and adsorption properties.

A compelling demonstration of SAW sensing performance was provided by Wang et al.[132], who developed a $H_2S$ sensor using $SnO_2$/CuO composite films on 36° YX $LiTaO_3$ delay lines ($f_r = 147\,MHz$). Their sensor achieved sharp switch-like responses within 55/45 $s$ over $2 - 12\,ppm$ ($2.79 - 16.73\,mg/m^3$) concentration ranges, with a high sensitivity of 16.9 $kHz/ppm$ (12.1 $kHz\,per\,mg/m^3$) at 160°C (Fig. 6c). Love-wave biosensors have also expanded into the bio-detection regime. Han et al.[133] reported $TiO_2$-guided devices that reached antigen detection limits as low as 0.7 $pg/mL$. Bahos et al.[134] developed a Love-wave e-nose coated with ZIF-8, ZIF-67, and Au nanoparticles capable of detecting VOCs in diabetic breath—acetone, ethanol, and ammonia—down to ~5 $ppm$ ($3.5 - 11.9\,mg/m^3$).

Continued improvements through nanostructuring and hybrid material integration have further enhanced SAW sensing performance. Li et al[135] developed SAW sensor with improved performance of $NO_2$ detection using a MXene/graphene oxide (GO) composite as a sensing material achieving high sensitivity of $NO_2$ ($\Delta f = 260\,Hz/ppm$ (138 $Hz\,mg/m^3$) across 1-200 ppm). Devkota et al.[18] improved reflective delay-line SAW sensitivity by integrating ZIF-8 MOF nanoparticles ($t = 240\,nm$). Operating at $f_r = 860\,MHz$ provided at least

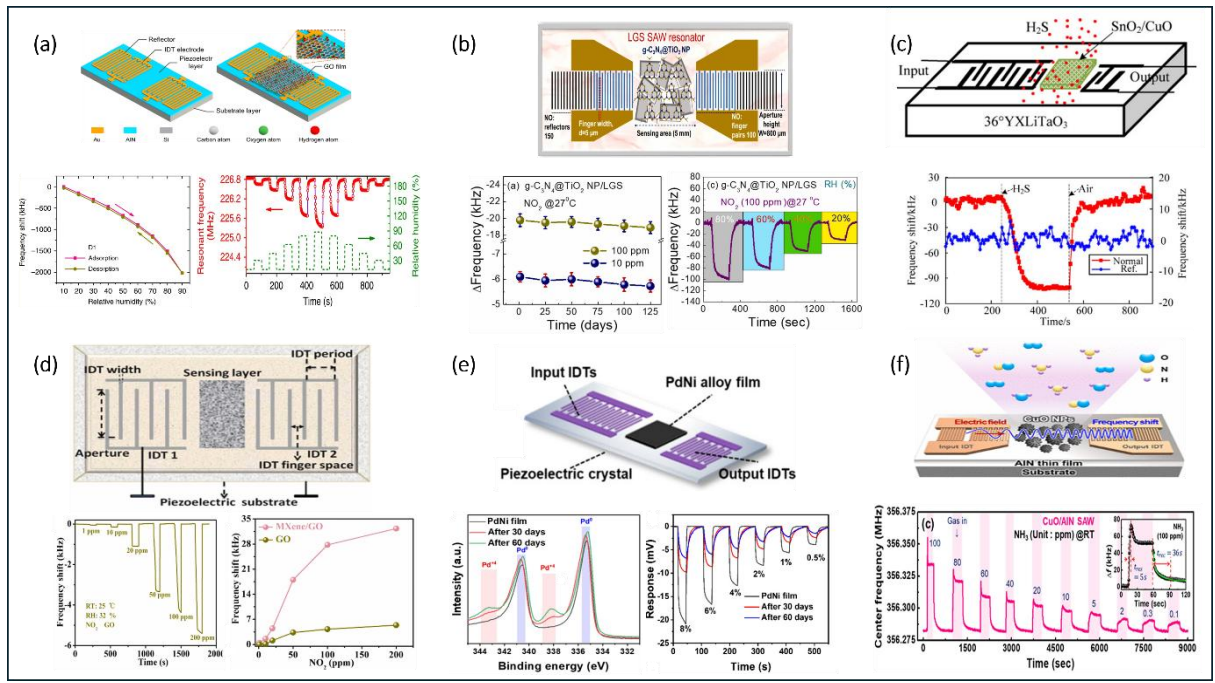

**Fig. 6.** Representative SAW-based gas sensors demonstrated in the literature. **(a)** Schematic of an AlN-based SAW humidity sensor functionalized with a graphene oxide (GO) film, exhibiting dynamic frequency response and humidity hysteresis behavior. The device operates at a resonant frequency of $226.3\ MHz$ with a GO thickness of $210\ nm$. *Reproduced with permission from*[104]. **(b)** Langasite (LGS) SAW resonator coated with a g-$C_3N_4$@$TiO_2$ nanocomposite, showing dynamic frequency shifts under 100 ppm $NO_2$ at various relative humidity levels (20%, 40%, 60%, 80%) and long-term stability for 10 and 100 ppm $NO_2$ at room temperature. *Reproduced with permission from*[131]. **(c)** Schematic of a $SnO_2$/CuO bilayer-coated SAW sensor designed for $H_2S$ detection, including comparative frequency response from active and reference sensors at 160 °C. *Reproduced with permission from*[132]. **(d)** MXene-activated oxide enhancing $NO_2$ capture showing dynamic $NO_2$ gas sensing frequency responses with GO and comparing with Mxene/GO and GO. *Reproduced with permission from*[135]. **(e)** SAW-based hydrogen sensor employing a PdNi sensing layer, with XPS spectra of Pd 3d core levels for fresh and aged films (30 and 60 days), along with corresponding response to $H_2$ at varying concentrations. *Reproduced with permission from*[136]. **(f)** Schematic of an AlN/CuO heterostructure SAW sensor showing dynamic frequency shifts in response to $NH_3$ concentrations ranging from 100 ppm down to 0.1 ppm at room temperature. *Reproduced with permission from*[56].

4× higher sensitivity than $430\ MHz$ devices for $CO_2$ and $CH_4$, consistent with Rayleigh wave mass sensitivity scaling. Jin et al.[136] developed PdNi alloy-based SAW sensors and showed that modulation of Pd oxidation states via PdHx formation improved $H_2$ sensitivity significantly, achieving a $6.4\ s$ response time and $2.23\ mV/\%$ sensitivity (Fig. 6e). Similarly, Bak et al. demonstrated CuO/AlN heterostructure SAW sensors operating at $f_r = 356.282\ MHz$ for $NH_3$ detection ($100 - 0.1\ ppm$), achieving a detection limit of $\sim 24\ ppb$ ($16.7\ \mu g/m^3$), $0.52\ kHz/ppm$ ($0.747\ Hz$ per $\mu g/m^3$) sensitivity, and ultra-fast $5\ s/25\ s$ response/recovery times (Fig. 6f). Table 6 provides a consolidated comparison of SAW gas sensors, detailing the target analyte, sensing layer, resonant frequency ($f_r$), sensitivity, response/recovery times, and operating temperature.

**Table 6:** A comprehensive summary of SAW-based gas sensors reported in the literature, highlighting key sensor parameters including target gas, sensing material, resonant frequency $f_r$, sensitivity, response and recovery times, and operating temperature. The table serves to illustrate performance trends and material choices across a wide range of gas-sensing applications. (Note: RT = Room Temperature*.)

| Target Gas | Sensing Layer | $f_r$ (MHz) | Sensitivity | | Response Time (s) | Recovery Time (s) | Temperature (°C) | Ref. |
|---|---|---|---|---|---|---|---|---|
| $H_2$ | $InO_x$ (RF diode sputtering) | 124 | 11.83 kHz/400 ppm | 0.359 kHz/mgm$^{-3}$ | - | 5 (min) | RT | [19] |
| $CH_4$ | Cryptophane-E | 204.2 | 118.8 Hz/% | - | 50 | 72 | RT | [20] |
| $NO_2$ | $Bi_2S_3$ Nanobelts | 200 | 2 kHz/10 ppm | 0.106 kHz/mgm$^{-3}$ | - | - | RT | [52] |
| $H_2S$ | $ZnO-Al_2O_3$ Nanocomposite | 200 | 500 Hz/10 ppb | 35.87 Hz/μgm$^{-3}$ | - | - | RT | [53] |
| $NO_2$ | GO-PEDOT:PSS | 136.10 | 5.89 kHz/100 ppm | 0.031 kHz/mgm$^{-3}$ | 35 | 10 | RT | [57] |
| $H_2$ | Cu-doped $SnO_2$ | 284 | 0.069 kHz/ppm | 0.837 kHz/mgm$^{-3}$ | 17 | 27 | RT | [113] |
| $SO_2$ | AgNWs@$SnO_2$/CuO | 696.02 | 3.696 kHz/ppm | 1.411 kHz/mgm$^{-3}$ | 40 | 60 | 80 | [114] |
| DMMP | Porous Graphene/PVDF | 434.9 | 1.407 kHz/ppm | 0.277 kHz/mgm$^{-3}$ | 4.5 | 5.8 | RT | [126] |
| $NO_2$ | Indium Tin Oxide | 440 | - | - | 90 | 15 | 240 | [127] |
| $CO_2$ | Teflon AF 2400 | 440 | - | - | 50 | 20 | RT | [127] |
| Humidity | Fluorinated Polyimide (PI) | - | 4.15 kHz/RH | - | 7 | 13 | RT | [128] |
| HCl | Cellulose Nanocrystals (CNCs) | 200 | 2 kHz/ppm | 1.341 kHz/mgm$^{-3}$ | 45 | 500 | RT | [129] |
| Humidity | Graphene Oxide (GO) | 226.3 | 111.7 ppm/%RH | - | 10 | 9 | RT | [104] |
| CO | Ppy@PEDOT:PSS | 136.41 | 9.38 kHz/ppm | 8.188 kHz/mgm$^{-3}$ | 121 | 138 | RT | [137] |
| VOCs | Calix[n]arene on AuNRs/AgNCs | 433.8 | - | - | 21 | 21 | RT | [138] |
| VOCs | $Fe_3O_4$ Nanoparticles | 69 | 3.1 kHz | - | 9 | - | RT | [139] |
| $NH_3$ | ZnO Nanorods | 199.95 | 1095 Hz/ppm | 1,572 Hz/mgm$^{-3}$ | 151 | 568 | RT | [140] |
| $NH_3$ | GO-$SnO_2$ Composite | 200 | 0.098 mV/ppb | 0.141 mV/μgm$^{-3}$ | 16.4 | 16.6 | RT | [141] |
| DMMP | Cu-$ZrO_2$ Nanocomposite | 4980 | 1198.2 kHz/ppm | - | 39 | 25 | RT | [142] |
| $SF_6$ | Multi-wall CNTs (MWCNTs) | 203.5 | 7.4 kHz/ppm | 1.239 kHz/mgm$^{-3}$ | - | - | RT | [143] |
| $O_2$ | $Zn_xFe_yO$ Films | 335 | - | - | 159 | - | RT | [144] |
| Acetone | Suspended Graphene/$SiO_2$ | 123 | 500 ppt | - | <125 s | < 75 s | RT | [106] |
| $NO_2$ | 2-D g-$C_3N_4$@$TiO_2$ | 135.73 | 19.7 kHz/ 100 ppm | 0.105 kHz/mgm$^{-3}$ | 143 | 114 | RT | [131] |
| $H_2S$ | $SnO_2$/CuO | 147 | 16.9 kHz/ppm | 12.125 kHz/mgm$^{-3}$ | 55 | 45 | 160 | [132] |
| $NH_3$ | CuO/AlN | 356.282 | 52.60 kHz/ 100 ppm | 0.755 kHz/mgm$^{-3}$ | 5 | 25 | RT | [56] |
| TNT | GO/PDEA/AuNR/PATP | 448.4 | 70.67 kHz/ 80 ppm | 0.095 kHz/mgm$^{-3}$ | 6.2 | 11.8 | RT | [145] |
| $NO_2$ | 3D- rGO-PPy/Ag (UV activation) | 246.99 | 127.68 Hz/ppm | 67.86 Hz/mgm$^{-3}$ | 36.7 | 58.5 | RT | [146] |
| DMMP | SXFA polymer | 433 | 23.5°/ppm | 4.63°/(mg/m³) | - | - | RT | [147] |
| $NO_2$ | SnS-CQD | 200 | 0.18 kHz/ppm | 0.096 kHz/mgm$^{-3}$ | 180 | 466 | RT | [148] |
| $NH_3$ | ZnO@MXene (UV Assisted) | 161.307 | 32.24 kHz/20 ppm | 2.314 kHz/mgm$^{-3}$ | 92 | 104 | RT | [125] |

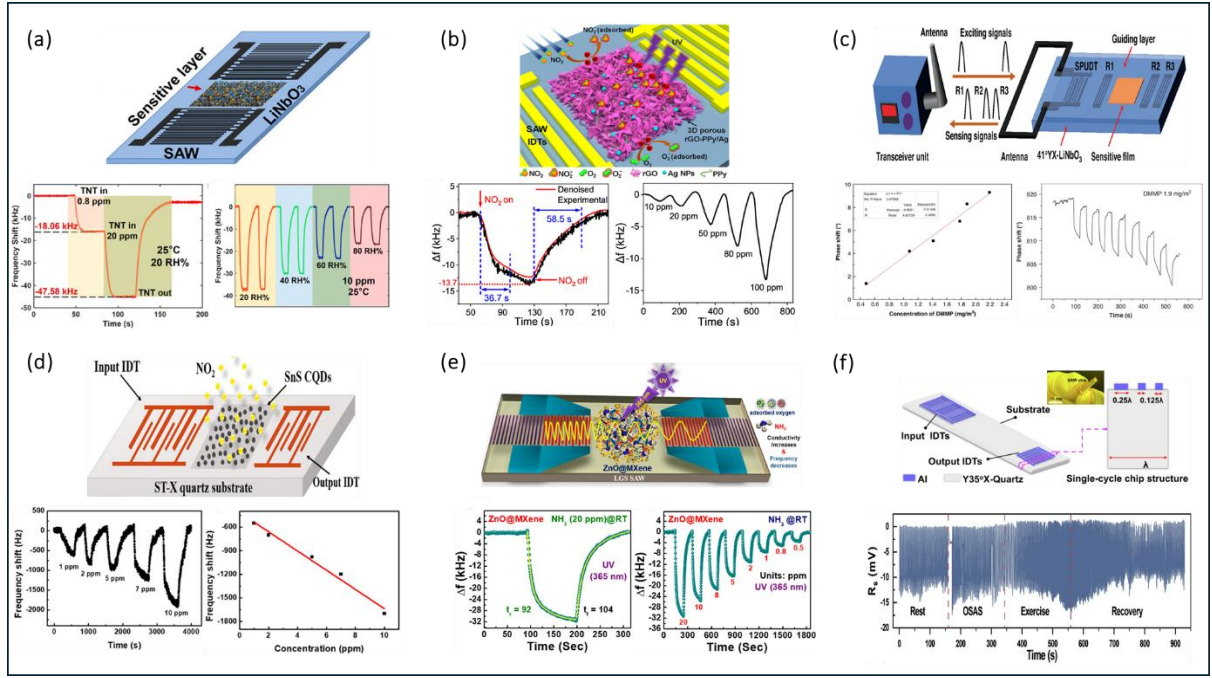

**Fig. 7.** Representative SAW-based gas sensors reported in the literature, demonstrating varied sensing mechanisms, materials, and configurations. **(a)** Schematic of a GO/PDEA/AuNR/PATP nanocomposite-based SAW sensor along with the frequency response to 0.8 and 20 ppm of TNT, and dynamic response curves for 10 ppm TNT under different humidity levels at 25 °C. *Reproduced with permission from*[145] **(b)** Schematic illustration of the $NO_2$ adsorption mechanism on a UV-activated 3D rGO-PPy/Ag SAW sensor, showing response/recovery time and dynamic response curves for $NO_2$ concentrations from 10 to 100 ppm. *Reproduced with permission from*[146] **(c)** Schematic and working principle of a 433 MHz WSAW chemical sensor for DMMP detection, including the response signal vs. DMMP concentration (21 °C, 33% RH) and repeatability over eight cycles. *Reproduced with permission from*[147] **(d)** Schematic of a SnS-CQDs SAW sensor, showing response curves across varying $NO_2$ concentrations and the corresponding concentration-dependent shifts. *Reproduced with permission from*[149] **(e)** Schematic of a ZnO@MXene hybrid heterostructure SAW sensor for $NH_3$ detection under UV illumination (365 nm @ 6.56 mW/cm²), including dynamic response curves at room temperature for concentrations ranging from 0.5 to 20 ppm. *Reproduced with permission from*[125] **(f)** Structural diagram of a Y35°X Quartz SAW chip employing acoustic impedance effects, including real-time sensor responses under simulated rest, OSAS, exercise, and post-exercise recovery breathing conditions. *Reproduced with permission from*[150].

Yang et al.[145] developed a $448.4\ MHz$ SAW sensor based on a GO/PDEA/AuNR/PATP nanocomposite for TNT detection. The sensor achieved detection down to $40\ ppb$ (372 µ$g/m^3$), with a sensitivity of $\sim 0.88\ kHz/ppm$ (0.0947 $Hz/\mu gm^{-3}$ and rapid $6.2/11.8\ s$ response/recovery (Fig. 7a). Its selectivity stemmed from multiple interactions with TNT molecules, including Meisenheimer complexation, $\pi - \pi$ stacking, hydrogen bonding, and deprotonation-assisted resonance stabilization. Xiong et al.[146] fabricated a UV-activated 3D rGO–PPy/Ag aerogel-based SAW sensor for $NO_2$ detection, achieving $127.68\ Hz/ppm$ (67.85 $Hz$ per $mg/m^3$) sensitivity with $36.7/58.5\ s$ response/recovery at room temperature. Performance enhancements were attributed to the porous architecture, photocarrier generation, and p–p heterojunctions (Fig. 7b). A $433\ MHz$ wireless SAW gas sensor was demonstrated by Pan et al. using a viscoelastic SXFA polymer layer for DMMP detection. The sensor showed a sensitivity of $4.63°/(mg/m^3)$, detection limit of $0.48\ mg/m^3$, and reliable operation across

–30 to 100 °C, though high humidity (>60% RH) impacted performance stability (Fig. 7c). Li et al.[149] introduced a SAW device ($f_r = 200\ MHz$) using spin-coated SnS colloidal quantum dots (CQDs), enabling room-temperature NO₂ detection with $0.18\ kHz/ppm$ ($95.6\ Hz\ per\ mg/m^3$) sensitivity and a theoretical detection limit of $\sim 52\ ppb$ ($97.9\ \mu g/m^3$). This performance was driven by high NO₂ adsorption energy on SnS surfaces (Fig. 7d).

Table 7. Representative sensing materials employed in SAW-based gas sensors, detailing their target analytes, dominant sensing mechanisms, key attributes, and references. Materials include nanostructures, polymers, and hybrids designed to enhance sensitivity, selectivity, and multimodal signal transduction.

| Sensing Material | Target Analyte | Dominant Sensing Mechanism | Key Attributes | Ref |
|---|---|---|---|---|
| Cryptophane-E | $CH_4$ | Mass loading | High adsorption capacity for small molecules | [20] |
| AgNWs@SnO2/CuO nanocomposites | $SO_2$ | Mass + Elastic loading | Enhanced sensitivity via heterojunction effects | [114] |
| Ppy@PEDOT:PSS | CO | Mass loading | Good flexibility and high carrier mobility | [137] |
| GO-PEDOT:PSS | $NO_2$ | Mass + Acoustoelastic effect | High surface area and tunable conductivity | [57] |
| Indium Tin Oxide (ITO) | $NO_2$ | Mass loading | High conductivity, used in dual-gas sensors | [127] |
| Bi₂S₃ Nanobelts | $NO_2$ | Mass loading | 1D nanostructure improves adsorption kinetics | [52] |
| Teflon AF 2400 | $CO_2$ | Mass loading | Chemically inert, selective for $CO_2$ | [127] |
| Graphene Oxide (GO) | Humidity | Mass + Elastic loading | Hydrophilic functional groups | [104] |
| Fluorinated Polyimide | Humidity | Mass loading | Hydrophobicity control for humidity detection | [128] |
| InOx thin films | $H_2$ | Mass + Elastic loading | High oxygen vacancy concentration | [19] |
| Cu-doped SnO₂ | $H_2$ | Mass + Acoustoelastic effect | Enhanced electron transfer with dopants | [113] |
| Calix[n]arene-functionalized AuNRs/AgNCs | VOCs | Mass loading | High affinity to VOC molecules | [138] |
| Fe₃O₄ nanoparticles | VOCs | Mass loading | Porous structure for fast adsorption/desorption | [139] |
| Cellulose Nanocrystals (CNCs) | HCl | Mass + Elastic loading | High surface area and moisture sensitivity | [129] |
| ZnO Nanorods | $NH_3$ | Mass loading | High carrier mobility and adsorption surface | [140] |
| GO-SnO₂ nanocomposite | $NH_3$ | Mass + Acoustoelastic effect | Improved sensitivity and selectivity via hybridization | [141] |
| ZnO-Al₂O₃ nanocomposite | $H_2S$ | Elastic loading dominant | Formation of ZnS increases elastic modulus | [53] |
| Porous Graphene/PVDF | DMMP | Mass loading | Porous structure for enhanced analyte trapping | [126] |
| Cu-ZrO₂ nanocomposite | DMMP | Mass loading | High catalytic activity and stability | [142] |
| MWCNT | $SF_6$ | Mass loading | Large surface area for gas adsorption | [143] |

| Nanostructured $Zn_xFe_yO$ films | $O_2$ | Mass loading | Oxygen adsorption modifies electrical properties | [144] |
| Suspended Graphene/$SiO_2$ | Acetone | Mass + Elastic loading | Minimal damping, high surface area | [106] |

A recent innovation by Pasupuleti et al.[125] demonstrated a UV-assisted SAW sensor based on a ZnO@MXene Schottky heterostructure ($f_r = 161.307\ MHz$), achieving sub-ppb NH$_3$ detection at room temperature. The system reported a $\sim 32.24\ kHz/20\ ppm$ ($2.31\ kHz/mgm^{-3}$), $92/104\ s$ response/recovery times, and a detection limit as low as $89.41\ ppb$ ($62.3\ \mu g/m^3$), facilitated by enhanced charge transfer and band bending via oxygen vacancies at the interface (Fig. 7e). In a distinct implementation, Cui et al.[150] developed a Y35X-quartz-based micro-nano SAW sensor ($f_r = 200\ MHz$) leveraging acoustic impedance modulation. This device achieved sub-second ($< 1\ s$) detection across a 0–100 v/v% range, with a detection limit of $\sim 1$ v/v%. It proved highly effective for real-time respiratory monitoring, capturing nuanced changes in breathing frequency ($f_{max} \approx 0.32 - 0.36\ Hz$), depth, and phase stability (Fig. 7f). Table 7 compiles SAW sensing materials as employed across diverse studies, linking each analyte to its dominant transduction route and to the design features most often exploited, surface functionalization, porosity, and heterojunction engineering.

While the judicious selection and nanostructuring of sensing materials have substantially elevated the performance of SAW gas sensors, material-driven gains alone do not fully capture the complexity of signal behavior under real-world, low-concentration, or transient conditions. Increasingly, it is becoming evident that the temporal and spectral evolution of the SAW sensor response especially in the presence of stochastic adsorption-desorption kinetics[124], holds critical information beyond traditional steady-state metrics. By embracing time-domain and frequency-resolved analysis, particularly within the framework of fluctuational kinetics, new opportunities emerge to probe ultra-low concentration regimes, discriminate overlapping analyte signals, and achieve faster, more adaptive sensing. The following section explores this emerging paradigm, highlighting its physical basis, analytical advantages, and application scope in SAW sensor technologies.

### 4.2. Time-Domain and Frequency-Resolved Performance of SAW Sensors: Advantages and Application Scope

Surface Acoustic Wave (SAW) sensors possess unique capabilities derived from their broad tunable frequency range—spanning from MHz to several GHz—allowing customization of sensor performance across a diverse set of environmental and functional conditions. Time-domain SAW configurations, such as delay-line devices, measure shifts in time delay or phase as acoustic waves propagate along a piezoelectric substrate. These variations capture dynamic gas–surface interactions and offer real-time response analysis. Such temporal signals are critical for passive wireless sensing, which operates via RF backscattering and eliminates the need for local power. This is especially valuable in embedded or inaccessible environments including implantable biomedical sensors, structural health monitoring, and distributed environmental networks[147].

In the frequency domain, SAW sensor sensitivity scales approximately as $\Delta f \propto f^2$. This quadratic dependence implies that higher-frequency devices ($> 500\ MHz$) exhibit significantly enhanced mass sensitivity and can detect minute perturbations like sub-ppb gas adsorption or changes in viscoelastic properties. This has enabled ultra-trace detection levels, extending into parts-per-trillion concentrations[106,151]. However, these gains often come with trade-offs—such as higher insertion losses, increased temperature sensitivity, and potential signal degradation in lossy environments. By contrast, lower-frequency devices ($50 - 300\ MHz$) provide greater thermal robustness, lower power consumption, and better signal-to-noise performance under challenging field conditions, making them suitable for wearables and remote deployments. Compared to conventional chemical transducers like Quartz Crystal Microbalance (QCM) or chemiresistive sensors, SAW sensors stand out due to multi-mode perturbation sensing, high spatial resolution, and wireless readout. QCMs, while gravimetrically precise, operate at low frequencies ($\sim 5 - 10\ MHz$) and suffer from limited miniaturizability. Chemiresistive sensors, although compact and inexpensive, struggle with baseline drift, poor selectivity, and humidity interference. SAW sensors, on the other hand, can simultaneously capture mass loading, viscoelastic changes, and acoustoelastic effects in real-time.

Importantly, SAW devices offer an unprecedented opportunity to probe non-equilibrium phenomena at nanoporous solid–gas interfaces. Materials such as MOFs, COFs, and hierarchical oxides exhibit stochastic adsorption dynamics governed by fluctuational kinetics—short-lived, metastable binding events not captured by conventional Arrhenius or equilibrium-based models. These adsorption-desorption events manifest as microsecond- to millisecond-scale fluctuations in sensor output, which encode analyte-specific kinetic signatures. SAW sensors, with their nanosecond-to-millisecond resolution and high sensitivity, are uniquely positioned to monitor these fluctuational signatures in real time. Studies using ultrathin MOF films (e.g., MFU-4) grown directly on SAW delay lines have demonstrated sub-ppm detection limits and millisecond-scale responses to gases[152]. As such, non-equilibrium thermodynamics becomes a powerful framework: it allows interpreting gas–surface interactions in terms of entropy production, metastability, and stochastic energy dissipation. By monitoring adsorption-relaxation trajectories through high-resolution SAW signals, researchers can move beyond average sensor outputs to access molecular fingerprints hidden in transient fluctuations[124].

This fluctuation-informed sensing approach is poised to revolutionize selectivity in SAW sensors. By capturing adsorption kinetics rather than just equilibrium endpoints, future SAW systems can speciate molecules, discriminate between similar analytes, and enable real-time kinetic profiling—paving the way for novel gas sensing paradigms in environmental monitoring, breath diagnostics, and industrial process control.

## 5. Machine learning for SAW sensors

As gas sensing applications become increasingly complex, the integration of machine learning (ML) techniques with surface acoustic wave (SAW)-based gas sensors has emerged as a powerful strategy to enhance performance, robustness, and adaptability. SAW sensors generate intricate signal outputs that are highly sensitive not only to target gas interactions but also to external factors such as temperature, humidity, UV radiation, baseline drift from material aging, cross-sensitivity to chemically similar gases, and signal nonlinearity. These overlapping influences often degrade long-term sensor stability and reliability. Traditional analytical approaches frequently struggle to decouple or interpret these interdependent effects accurately. Machine learning offers a data-driven alternative capable of learning nuanced patterns in sensor behaviour, improving gas identification, concentration estimation, and real-time decision-making—ultimately paving the way for more intelligent and reliable SAW gas sensing systems.

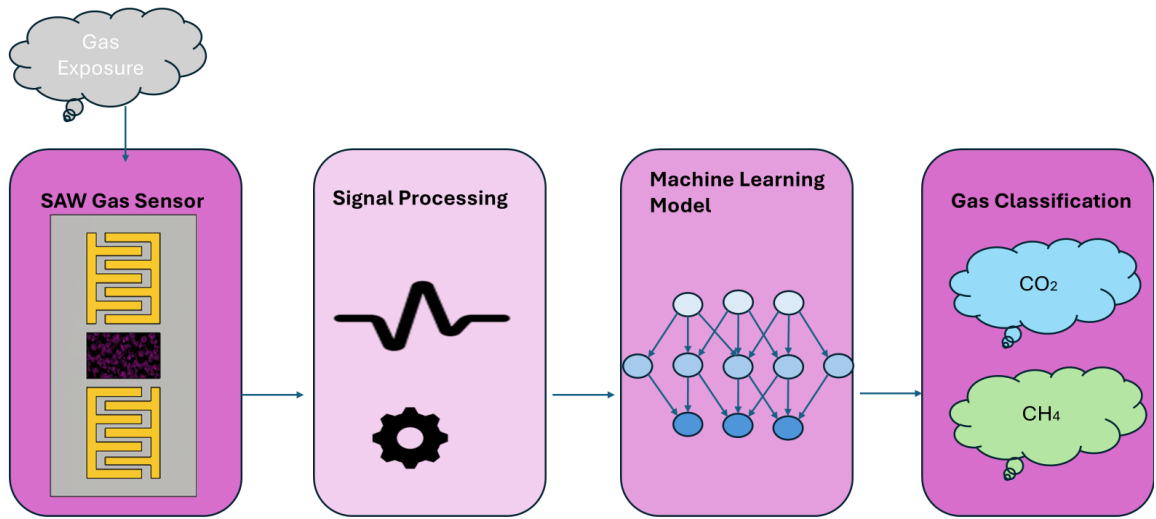

**Fig. 8.** Machine learning pipeline for SAW-based gas classification. Gas exposure alters the sensor's acoustic signal, which is processed and input into a trained ML model for accurate classification of analytes such as $CO_2$ and $CH_4$.

Fig. 8 depicts a typical ML pipeline for SAW gas sensors. It begins with acquisition of raw sensor data, followed by preprocessing operations such as denoising and normalization to improve signal quality and consistency. The processed data is then fed into supervised ML models trained for gas classification and concentration prediction. Depending on task complexity and computational constraints, techniques range from classical algorithms like random forests to deep learning architectures like artificial neural networks (ANNs).

Although SAW sensors exhibit excellent sensitivity, they may struggle to differentiate gases with similar physicochemical interactions, leading to poor selectivity. Many gases can induce comparable shifts in the frequency response, complicating interpretation using conventional signal processing techniques. ML models trained on labelled datasets have demonstrated the ability to overcome these challenges, accurately classifying multiple gases even in the presence of strong cross-sensitivity. For instance, Singh et al. (2016)[153] applied a combined approach where Principal Component Analysis (PCA) was used to extract features from the frequency response of a polymer-coated SAW sensor. These features were then input into an ANN to classify four toxic analytes across broad concentration and temperature ranges with high accuracy. Similarly, Zhou and Liu (2021)[154,155] proposed a deep learning model based on a convolutional long short-term memory (CLSTM) architecture. The model fused convolutional layers (for spatial feature extraction) with LSTM layers (for capturing temporal dependencies), enabling real-time discrimination of CO, $CH_4$, and $C_3H_8$ at various concentrations. The eight-layer CLSTM model significantly outperformed conventional methods such as support vector machines.

Beyond gas classification, ML techniques have proven effective in compensating for environmental interferences such as temperature, humidity, and UV radiation. Tan et al.[156] tackled this by extracting 11 signal features from SAW sensor outputs and applying a dual-model strategy: a Support Vector Machine (SVM) for temperature prediction and a Random Forest Regressor (RFR) for estimating UV intensity. The approach achieved high accuracy, with models selected based on task-specific performance. Building on this, Xia et al. (2025)[157] expanded the framework to handle simultaneous interference from humidity, temperature, and UV radiation. They assessed a diverse set of models, including linear regressors, decision trees, and nonlinear methods such as XGBoost. While temperature and humidity predictions were well handled by tree-based models, UV response proved more complex.

To address this, they implemented a stacking ensemble technique, where a meta-model was trained on outputs from base learners. This significantly reduced UV prediction error and improved model robustness under real-world conditions.

To counter long-term sensor drift and environmental variability, ensemble learning with periodic retraining has emerged as a practical solution. Lai et al. (2022)[155] evaluated four recurrent neural network (RNN) architectures—LSTM, GRU, Bi-LSTM, and Bi-GRU—for predicting concentrations of $CO$, $O_3$, and $NO_2$ from raw time-series data. Outputs from these models were combined using a fully connected neural network serving as a meta-learner. Moreover, the authors introduced a dynamic retraining strategy that updates model weights incrementally, without requiring full retraining. This continuous learning capability enables the system to adapt to evolving signal characteristics, improving resilience and extending sensor lifespan in low-power embedded platforms.

Despite these advancements, key challenges persist in real-world deployment. Most reported studies are conducted in controlled laboratory settings, where analyte types and environmental parameters are manually selected. This limits dataset diversity and can impair generalizability to field conditions, where sensor exposure is uncontrolled and noisy. The deployment of ML models on edge devices also presents significant hurdles. Resource-constrained environments may lack the computational power to support complex models. Here, techniques such as model compression, pruning, quantization, and knowledge distillation become essential to reduce model size and latency without sacrificing performance. Designing lightweight ML architectures compatible with edge hardware is particularly important for real-time analysis in passive SAW sensors, which often operate without continuous power. An equally critical issue is the explainability of ML models. While black-box models can deliver strong performance, their lack of interpretability is problematic in high-stakes domains like healthcare, industrial safety, and environmental compliance. Adopting interpretable models and leveraging explainability tools—such as SHAP values or physics-informed AI—can enhance transparency and user trust.

These developments signal a transformative step toward intelligent and adaptive gas sensing systems, leveraging ML to extract deeper insights and enhance SAW sensor resilience in dynamic environments. These hybrid systems hold great promise for next-generation sensing platforms aligned with IoT and edge computing paradigms. Nevertheless, advancing reliable real-world deployment will depend on addressing data scarcity, model explainability, hardware constraints, and dynamic environmental variability.

## 6. Recommendations for future development of SAW gas sensors

Despite the rapid evolution of SAW sensors in gas detection, several technical barriers remain that must be overcome to ensure robust, scalable deployment across real-world environments.

**Sensor film engineering** remains a foundational issue. Gas adsorption not only alters surface mass but can dynamically affect mechanical properties, such as elasticity and stiffness, which modulate acoustic wave propagation. Future studies must go beyond mass-loading paradigms and explore how these dynamic mechanical shifts impact signal response and how they can be exploited for additional sensing modes.

**Selectivity** remains a central challenge, particularly in the presence of chemically similar analytes. While previous sections highlighted MOFs, COFs, and advanced signal processing via ML to improve discrimination, future progress hinges on designing multi-functional interfaces that integrate selective adsorption, catalytic transformation, and reaction kinetics tailoring. Combinatorial coatings and heterostructured interfaces may offer new routes to encode richer chemical information into acoustic signals.

**Environmental robustness** also warrants deeper attention. Prior sections discussed temperature and humidity compensation via machine learning; however, materials-based solutions—such as temperature-insensitive cuts, composite substrates, and thermally stable coatings—need to be pursued in tandem with software solutions for redundancy and reliability[106]. Hybrid sensor modules that fuse physical redundancy with AI-based prediction can form the backbone of robust SAW systems in uncontrolled field deployments.

**Signal quality**, particularly at high frequencies, is susceptible to insertion loss, noise, and modal dispersion. As previous sections introduced the benefit of GHz-scale devices for enhanced sensitivity, here it is must to reiterate that achieving this requires next-generation materials and microfabrication strategies—including low-loss piezoelectric films, SPUDT architectures, and finely pitched IDTs manufactured via advanced lithographic techniques.

**Durability and longevity** are key to sustainable deployment, especially in harsh environments such as industrial exhausts, biofluids, or high-humidity conditions. Research into self-healing polymers, corrosion-resistant coatings, and encapsulation methods is vital to ensure long-term operational stability. Strategies that maintain acoustic transparency while protecting sensitive layers are crucial.

**Stability and reproducibility** together define the ability of SAW sensors to maintain consistent performance under varying environmental conditions and to deliver identical responses across repeated trials. Optimized IDT configurations and sensing layers with low susceptibility to humidity and temperature fluctuations can significantly enhance these characteristics.

**Miniaturization and high-frequency scalability** introduce unique challenges. As SAW sensors push into the GHz regime to achieve ultra-high mass sensitivity, this direction demands advanced lithographic capabilities for fabricating ultra-fine-pitch IDTs, along with resolving issues like acoustic wave dispersion and thermal instability at nanoscale dimensions. Strategies to mitigate these include the use of high-velocity piezoelectric materials such as ScAlN and $LiNbO_3$ variants, coupled with advanced packaging techniques to maintain performance integrity. Additionally, rotating IDT configurations for angle optimization and apodized/dummy electrode designs can help suppress spurious modes, improving spectral fidelity and overall sensor precision.

**Edge computing integration** is another future-critical domain. While ML methods enable real-time gas classification and environmental compensation, deploying these models on low-power SAW platforms requires model optimization (e.g., quantization, pruning, distillation) and hardware-aware ML design. On-chip learning and explainability remain underdeveloped areas for embedded SAW-ML systems.

**Sustainability and eco-design** also deserve emphasis. The development of biodegradable electrodes, recyclable substrates, and energy-harvesting interfaces aligns SAW sensors with global efforts in green electronics. Moreover, advanced IDT configurations—such as apodization and dummy fingers—can suppress spurious modes, enhancing spectral clarity and sensor accuracy.

**Networked intelligence** is the future horizon. SAW sensors integrated with IoT ecosystems, passive wireless interrogation, and real-time ML backends can enable autonomous, self-calibrating systems for urban air monitoring, industrial leak detection, and personalized health diagnostics. These systems will depend on breakthroughs not just in sensor physics, but also in ultra-low-power RF circuits, secure data protocols, and cloud-edge hybrid architectures.

In summary, realizing the full potential of SAW gas sensors demands coordinated innovation in:

- Material design
- Environmental compensation
- Microfabrication
- AI-assisted signal processing
- Mechanical integration
- Networked deployment

By resolving these interlinked challenges, SAW technology is poised to evolve into a versatile, intelligent, and sustainable platform for next-generation sensing across industrial, biomedical, and environmental domains.

# 7. Conclusion

SAW gas sensors have evolved into one of the most promising platforms for high-performance gas sensing, combining the advantages of fast response, miniaturization, wireless operability, and multimodal sensitivity. This review has presented a comprehensive landscape of the diverse material innovations including porous MOFs, conductive polymers, nanostructured oxides, and hybrid nanocomposites that underpin the sensitivity and selectivity of SAW-based gas sensors. The dynamic interplay between mass loading, elastic perturbation, and acoustoelastic effects provides a versatile transduction framework that allows SAW sensors to function across a broad spectrum of gaseous analytes and environmental conditions.

Beyond traditional sensing paradigms, we emphasize a transformative perspective rooted in non-equilibrium thermodynamics and fluctuational kinetics at the nanoporous solid–gas interface. SAW sensors, by virtue of their temporal resolution and responsiveness, are uniquely suited to extract analyte-specific signatures from transient adsorption–desorption events. This capability, long overlooked in ensemble-averaged models, opens avenues for real-time molecular speciation and kinetic fingerprinting—especially when supported by machine learning algorithms that decode complex, nonlinear sensor outputs.

The future of SAW gas sensors lies in the convergence of smart materials, advanced signal processing, and embedded intelligence. Integration with machine learning not only addresses key challenges like drift, environmental variability, and cross-sensitivity but also enables adaptive, self-correcting sensor networks. Emerging efforts to push SAW operation into the GHz regime, develop sustainable and flexible substrates, and exploit bioinspired design principles hold promise for truly ubiquitous, ultra-sensitive chemical sensing platforms.

To translate these scientific advances into real-world impact, coordinated efforts across material science, device engineering, data science, and system-level integration are essential. The next generation of SAW sensors will not only detect gases—they will interpret, adapt, and interact, forming the backbone of intelligent sensing infrastructures for health, environment, security, and beyond.

# Acknowledgements

This publication has emanated in part from research supported by a research grant from Science Foundation Ireland and the Department of Agriculture, Food and Marine on behalf of the Government of Ireland under the Grant 21/RC/10303_P2 (Vistamilk), and supported from research conducted with the financial support of Science Foundation Ireland (SFI) and is co-funded under the European Regional Development Fund under Grant Number 13/RC/2077_P2 (CONNECT)

## Author contributions

Conceptualization: SA, BS and DS
Visualization: SA, BS, DS and VS
Funding acquisition: VS, AOR and UR
Supervision: BS, VS, AOR and UR
Writing – original draft: SA, BS and DS
Writing – review & editing: SA, BS, DS, AOR, UR and VS

## Declaration of interests

The authors declare that they have no known competing financial interests or personal relationships that could have appeared to influence the work reported in this paper.

## Data and materials availability

All data is available in the main text. No new code was developed for this work.